\documentclass[prc,twocolumn,showkeys,showpacs,superscriptaddress,]{revtex4}

\usepackage{dcolumn}
\usepackage{graphicx,color}
\usepackage{textcomp}
\begin {document}
  \newcommand {\nc} {\newcommand}
  \nc {\beq} {\begin{eqnarray}}
  \nc {\eeq} {\nonumber \end{eqnarray}}
  \nc {\eeqn}[1] {\label {#1} \end{eqnarray}}
  \nc {\eol} {\nonumber \\}
  \nc {\eoln}[1] {\label {#1} \\}
  \nc {\ve} [1] {\mbox{\boldmath $#1$}}
  \nc {\ves} [1] {\mbox{\boldmath ${\scriptstyle #1}$}}
  \nc {\mrm} [1] {\mathrm{#1}}
  \nc {\half} {\mbox{$\frac{1}{2}$}}
  \nc {\thal} {\mbox{$\frac{3}{2}$}}
  \nc {\fial} {\mbox{$\frac{5}{2}$}}
  \nc {\la} {\mbox{$\langle$}}
  \nc {\ra} {\mbox{$\rangle$}}
  \nc {\etal} {\emph{et al.\ }}
  \nc {\eq} [1] {(\ref{#1})}
  \nc {\Eq} [1] {Eq.~(\ref{#1})}
  \nc {\Ref} [1] {Ref.~\cite{#1}}
  \nc {\Refc} [2] {Refs.~\cite[#1]{#2}}
  \nc {\Sec} [1] {Sec.~\ref{#1}}
  \nc {\chap} [1] {Chapter~\ref{#1}}
  \nc {\anx} [1] {Appendix~\ref{#1}}
  \nc {\tbl} [1] {Table~\ref{#1}}
  \nc {\fig} [1] {Fig.~\ref{#1}}
  \nc {\ex} [1] {$^{#1}$}
  \nc {\Sch} {Schr\"odinger }
  \nc {\flim} [2] {\mathop{\longrightarrow}\limits_{{#1}\rightarrow{#2}}}
  \nc {\textdegr}{$^{\circ}$}
  \nc {\IR} [1]{\textcolor{red}{#1}}
  \nc {\IB} [1]{\textcolor{blue}{#1}}
  \nc {\IG} [1]{\textcolor{green}{#1}}

\newcommand{\bea}{\begin{equation}}
\newcommand{\eea}{\end{equation}}

\title{Comparing non-perturbative models of the breakup of neutron-halo nuclei}
\author{P.~Capel}
\email{pierre.capel@centraliens.net}
\affiliation{National Superconducting Cyclotron Laboratory,
Michigan State University, East Lansing, Michigan 48824, USA}
\affiliation{Helmholtz-Insitut Mainz, Johannes Gutenberg-Universit\"at, D-55128 Mainz, Germany}
\affiliation{Physique Quantique C.P. 165/82, Universit\'e Libre de Bruxelles (ULB), B-1050 Brussels, Belgium}
\author{H.~Esbensen}
\email{esbensen@phy.anl.gov}
\affiliation{Physics Division, Argonne National Laboratory,
Argonne, Illinois 60439, USA}
\author{F.~M.~Nunes}
\email{nunes@nscl.msu.edu}
\affiliation{National Superconducting Cyclotron Laboratory
and Department of Physics and Astronomy,
Michigan State University, East Lansing, Michigan 48824, USA}
\date{\today}
\begin{abstract}
Breakup reactions of loosely-bound nuclei are often used to extract structure and/or astrophysical information.
Here we compare three non-perturbative reaction theories often used when analyzing breakup experiments, namely
the continuum discretized coupled channel model,
the time-dependent approach relying on a semiclassical approximation,
and the dynamical eikonal approximation.
Our test case consists of the breakup of \ex{15}C on Pb at 68~MeV/nucleon
and 20 MeV/nucleon.
\end{abstract}
\maketitle
\section{Introduction}

Due to the proximity to the particle threshold, loosely-bound nuclei
dissociate easily during collisions with nuclear targets.
Consequently, they are often studied through breakup reactions,
in which the loosely-bound particle(s) dissociates from the core
of the nucleus through interaction with the target.
In the following, we focus on \emph{elastic breakup}, i.e.,
a reaction in which the target is left in its ground state
and all projectile fragments are detected in coincidence after dissociation.
The use of breakup reactions for extracting properties of exotic
nuclei is numerous and varied, including one-neutron halo systems \cite{c19-riken,be11-gsi,c15-riken},
configuration-mixed systems \cite{o23-gsi,ne31-riken}, two-neutron halo systems \cite{he6-gsi,li11-riken,c22-riken},
as well as proton rich systems \cite{b8-gsi,f17-ornl,f17-catania}.
While the shape of the energy distribution can tell us about the separation energy and the angular
momentum of the ground state,  the magnitude of the cross section is related to the asymptotic
normalization of the ground state \cite{bu-anc}. In addition, for two-particle halo systems, one also
expects to obtain information on the correlations in the  valence pair \cite{correlation-2n}.
More recently, breakup reactions have proven to be a useful tool in exploring nuclei beyond
the dripline and studying decay modes of resonant states \cite{li12-nscl,be13-nscl,o24-nscl}.

If the reaction is dominated by the electromagnetic interaction, it is possible to connect the breakup cross section
with the capture cross section \cite{BBR86,BHT03}.
This method, known as the
Coulomb-dissociation method, is of interest to astrophysics because
it can provide radiative-capture cross sections at
very low relative energies where a direct measurement is not feasible.
It also gives access to neutron-capture cross sections by unstable
species, which are impossible to measure in the laboratory.
It has been applied to a number of cases
\cite{BHT03,b8-gsi,c15-riken,summers04,li6-gsi}.
Providing confidence that the Coulomb dissociation method works,
neutron capture cross sections for $^{14}$C($n$,$\gamma$)$^{15}$C were
extracted from the Coulomb dissociation data \cite{c15-riken} using two
independent methods \cite{summers08,esbensen09} and perfect agreement was
obtained when compared to direct measurements \cite{reifarth}.
Similar efforts have been performed for the breakup of $^8$B
\cite{solar,EBS05,GCB07,Oga06}.

The common feature of all the above mentioned
experiments is their need for a reliable reaction model in the analysis.
One needs to be careful with separating nuclear and Coulomb processes,
often nuclear-Coulomb interference is important and
the dynamical effects in the continuum are crucial \cite{nunes99,hussein06,EB02B,capel-cc,EBS05,BHT03}.
These results imply that, in general, {\em perturbative approaches} are not
accurate enough for a reliable analysis of Coulomb-breakup measurements.

Hand-in-hand with the experimental advances, a number of non-perturbative breakup theories have been developed improving
the method by which breakup reactions are studied (see \Ref{capel-review}
for a recent review). The many new developments rely on different approximations, have separate advantages and shortcomings,
and vary also in the level of complexity.
We believe it is timely to compare these theories and understand the level of accuracy of the approximations made.
In this work we compare the most common non-perturbative approaches to describe the breakup of a one-neutron halo nucleus that can be approximated
by a two-body cluster. These are:
i) the continuum discretized coupled channel
method (CDCC) \cite{cdcc,tostevin01}, which is fully quantal and does not make approximations in the projectile-target dynamics,
ii) the time-dependent approach (TD) \cite{kido-tdse,esbensen-tdse,typel-tdse,capel-tdse},
which is based on a semiclassical approximation \cite{AW75}
that describes the projectile-target relative motion by a classical trajectory, and
iii) the dynamical eikonal approximation (DEA) \cite{BCG05,capel-dea} which
relies on the eikonal approximation \cite{Glauber}.

All three theories are built on the same three-body description of the reaction:
the projectile $P$, described as a valence neutron $f$ loosely bound to
a core $c$, impinges on a target $T$ considered as inert.
The effective interaction between $c$ and $f$ is adjusted to reproduce known properties of the projectile, while
the interactions between the projectile fragments and the target are simulated
by optical potentials fitted to elastic-scattering data for the $c$-$T$ and $f$-$T$ systems.

For the type of reactions we are interested in here, the CDCC method is the most
accurate method available on the market but it is also the most computationally intensive
and requires elaborate model-space studies.
Developments that go beyond the inert-core and/or inert-target approaches, or extensions to $N$-body projectile
clusters ($N>2$) are compromised by computational limitations. On the opposite,
TD and DEA are not computationally intensive and
are rather straightforward to set up.
The question is whether these approximations can do a good job for the
reactions of interest. To answer this question one needs to quantify the level of accuracy of the
approximations introduced.

For a meaningful comparison, it is necessary that all three methods start from the exact
same three-body Hamiltonian.
Typical breakup observables are then compared to quantify the accuracy of the various approximations.
The test case chosen is $^{208}$Pb($^{15}$C, $^{14}$C $n$)$^{208}$Pb, a case where we expect the
three-body description to be adequate. Our study is performed at two energy regimes,
one at a typical energy in fragmentation facilities (68~MeV/nucleon) \cite{c15-riken}
and the other at the higher energy limit of ISOL facilities (20~MeV/nucleon).
We ignore for practical reasons the effect of relativity.

In \Sec{theory} we briefly summarize the three methods under scrutiny.
The model inputs are given in \Sec{inputs}.
In \Sec{results} the results for
breakup are presented, and conclusions are drawn in \Sec{conclusions}.
The details about the calculations in all three models can be found in the
Addendum provided as supplemental material.

\section{Brief theoretical description}
\label{theory}

\subsection{Common framework}
\begin{figure}[h]
\center
\includegraphics[width=6cm]{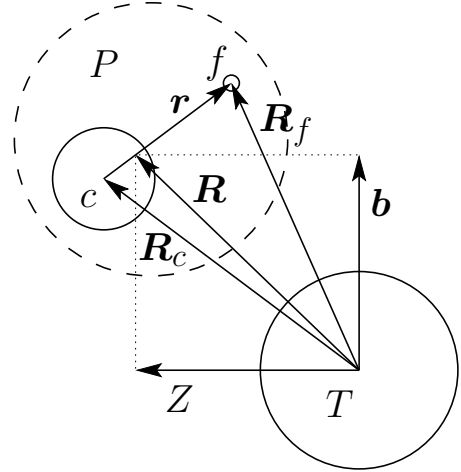}
\caption{Set of coordinates used in the reaction modeling.
The longitudinal $Z$ and transverse $\ve{b}$ components of
$\ve{R}$ are shown as well.}\label{f0}
\end{figure}

To study the breakup of a projectile $P$ into a core $c$ and
a valence neutron $f$ on a target $T$, we start from the (non-relativistic)
three-body Hamiltonian
\bea
H_{\rm 3b}(\ve{R},\ve{r})= \hat T_{\ve{R}} + H_0(\ve{r})
+ U_{cT}(\ve{R}_c) + U_{fT}(\ve{R}_f),
\label{h3b}
\eea
expressed in the set of coordinates illustrated in \fig{f0}.
In \Eq{h3b}, $\hat T_{\ve{R}}$ is the kinetic-energy
operator for the $P$-$T$ relative motion.
The two-body Hamiltonian $H_0$ describes the internal structure of the
projectile
\beq
H_0(\ve{r})=\hat T_{\ve{r}}+V_{cf}(\ve{r}),
\eeqn{h0}
where $\hat T_{\ve{r}}$ is the $c$-$f$ kinetic-energy operator and
$V_{cf}$ is an effective potential, modeling the $c$-$f$ interaction.
This potential is adjusted to reproduce the bound-state spectrum and low energy scattering
states of the projectile.
The optical potential $U_{cT}$ ($U_{fT}$) describes the elastic scattering
of the core (valence neutron) by the target and contains a Coulomb part and a nuclear part.

In all three methods, a partial wave expansion for the projectile states is used:
\bea
{\phi}_{k}^{ljIM}(\ve{r})=
\frac{u_{k}^{ljIM}(r)}{r}\, \left[\left[Y_l(\hat{\ve{r}})\otimes
{\cal X}_s \right]_j \otimes {\cal X}_{I_c} \right]_{I M} ,
\label{wf2}
\eea
where $Y$ is a spherical harmonic \cite{AS70} and $\cal X$s are spinors.
The quantum number $l$ is the orbital angular momentum of
$f$ relative to $c$, $s$($I_c$) is the spin of the fragment
$f$($c$), and the total angular momentum of the projectile is $I$
with projection $M$.
We denote by ${\phi}_0$ the projectile bound state of (negative) energy $E_0$.
For simplicity in this formulation we consider only one bound state.
In this manner, all the other eigenstates of $H_0$ correspond to positive energies
$E=\hbar^2/2\mu_{cf}k^2$, with $\mu_{cf}$ the $c$-$f$ reduced mass.
They describe the $c$-$f$ continuum.
Of course, the formulation can be easily extended to include bound excited states
and we do include these in the application presented in Section \ref{results}.

Within this framework, the study of the $P$-$T$ collision reduces to solving
the \Sch equation
\beq
H_{\rm 3b}\Psi(\ve{R},\ve{r})=E_{\rm tot}\Psi(\ve{R},\ve{r})
\eeqn{3beq}
with initial boundary condition:
\beq
\Psi(\ve{R},\ve{r})\flim{Z}{-\infty}e^{iK_0Z}\phi_0(\ve{r}),
\eeqn{boundary}
where the initial $P$-$T$ momentum $\hbar\ve{K}_0$ is assumed along the $Z$ axis.
Its norm is related to the total energy
$E_{\rm tot}=\hbar^2K_0^2/2\mu_{PT}+E_0$,
with $\mu_{PT}$ the $P$-$T$ reduced mass.
There are different assumptions used in the treatment of the full
three-body wave function $\Psi$.
We capture the essential features in the following subsections.

\subsection{Continuum discretized coupled channel method}\label{cdcc}

The full three body wave function can always be expanded in terms of the complete
set of projectile states $\phi_k^{ljIM}$ as:
\bea
\Psi(\ve{R},\ve{r})
= \phi_0(\ve{r}) \psi_{0}(\ve{R})+ \sum_{ljIM}\int  d k\  \phi_k^{ljIM}(\ve{r})
  \psi_K^{ljIM}(\ve{R}),
\label{cdccwf0}
\eea
such that the momentum $\hbar k$ between the internal motion of
$c+f$ is related to the momentum $\hbar K$ between the projectile center of mass
and the target through energy conservation
$E_{\rm tot}=\hbar^2K^2/2\mu_{PT}+\hbar^2k^2/2\mu_{cf}$.
An expansion involving an integral
over momentum is not tractable, so in CDCC a discretization of the projectile
continuum is performed \cite{book,cdcc}.
There are various ways of performing this discretization, and here we will use the
so-called average method whereby $\phi_k^{ljIM}$ is replaced by its average
over a momentum bin $[k_{p-1},k_p]$, $\widetilde \phi_p^{ljIM}$ \cite{book}. In this method the three-body wave function
is approximated by
\bea
\Psi^{\rm CDCC}(\ve{R},\ve{r}) = \sum_{ljIM}\sum_{p=0}^{N} \widetilde \phi_{p}^{ljIM}(\ve{r}) \psi_p^{ljIM}(\ve{R}),
\label{cdccwf}
\eea
with $p=0$ corresponding to the initial ground state and  $p \geq 1$ corresponding to the bin wave functions.
The sum runs up to $N$, which is associated to the maximum projectile excitation energy $E_{\rm max}$ considered
in the model space. In the end, the method needs to be independent of discretization and  model space,
and thus $E_{\rm max}$ needs to be large enough and the bin width needs to be small enough
to accurately describe the process of interest \cite{book}.

When introducing expansion \eq{cdccwf} into the full three-body equation \eq{3beq},
and after integrating over the angular variables and $r$, one arrives at the following
coupled-channel equations in $R$ \cite{book}:
\begin{eqnarray}
 \left [   -\frac{\hbar^2 }{2 \mu_{PT}}  \left ( \frac{d^2 }{ d R^2} - \frac{L(L{+}1)} {R^2} \right )
   +  E_p - E_{\rm tot} \right ]
 \chi_ {\alpha}^{J_{\rm tot}} (R)  \nonumber \\
 \hspace{2cm} +  \sum _ {\alpha'}
 i ^ {L' - L} ~ V _{\alpha \alpha'}^{J_{\rm tot}}(R)  \chi_{\alpha'}^{J_{\rm tot}}  (R)= 0\ ,
\label{cdcceq}
\end{eqnarray}
where
$L$ is the $P$-$T$ relative angular momentum,
$E_p$ is the midpoint energy of bin $p$,
and $\alpha$ is the index for the channel $\{pljIL\}$.
The coupling potentials $V _{\alpha  \alpha'}^{J_{\rm tot}}(R)$ are defined by,
\beq
\lefteqn{ V_{\alpha  \alpha'}^{J_{\rm tot}}(R)  =}\nonumber \\
& & \langle[\widetilde\phi_p^{ljI} Y_L(\hat{\ve{R}})]_{J_{\rm tot}} | U _ {cT} (\ve{R}_ c) + U _ {fT} (\ve{R}_f) |
[\widetilde\phi_{p'}^{l'j'I'} Y_{L'}(\hat{\ve{R}})]_{J_{\rm tot}}\rangle,
\eeqn{CP}
where $J_{\rm tot}$ is the total angular momentum resulting from
the coupling of $I$ and $L$.
Equation~\eq{cdcceq} is solved with scattering boundary conditions at large distances:
\bea
\chi_{\alpha}^{J_{\rm tot}}(R) \flim{R}{\infty} \frac{i}{2} \left[ H^-_{\alpha}(KR) \delta_{\alpha \alpha_i} - H^+_{\alpha}(KR) S_{\alpha \alpha_i}^{J_{\rm tot}}\right],
\eea
where $\alpha_i$ is the entrance channel,
and $H^\pm$ are Coulomb Hankel functions \cite{book}.
Breakup observables are then calculated from the resulting S matrix
\cite{tostevin01,fresco}.
In the present work, we use the code {\sc fresco} to numerically solve
the set of coupled equations \eq{cdcceq} \cite{fresco}.
The parameters of our calculations are given in the Addendum
provided as a supplemental material of this article.

\subsection{Time-dependent model}\label{tdse}
It can be very demanding to solve the coupled-channel equations \eq{cdcceq}
numerically. To reduce the computational cost, other models have been developed.
In the semiclassical approximation, the $P$-$T$ relative motion is
approximated by a classical trajectory $\ve{R}(t)$
\cite{kido-tdse,esbensen-tdse,typel-tdse,capel-tdse}.
Along that trajectory, the projectile experiences a time-dependent potential
that simulates its interaction with the target.
Assuming a quantal description of the internal structure of the projectile,
this approximation leads to the time-dependent \Sch equation,
\beq
i\hbar \frac{\partial}{\partial t}\Psi^{\rm TD}(t,b,\ve{r})=
\left[H_0+V_{PT}(t,\ve{r})\right]
\Psi^{\rm TD}(t,b,\ve{r}),
\eeqn{tdseeq}
where $b$ is the impact parameter characterizing the trajectory.
The time-dependent potential $V_{PT}$ appearing in this equation
is the sum of the optical potentials of the three-body Hamiltonian \eq{h3b},
from which the potential that generates the trajectory is subtracted \cite{esbensen-tdse}.

For Coulomb-dominated reactions, the potential that generates the
classical trajectory is usually the bare $P$-$T$
Coulomb interaction, i.e.,
\beq
V_{PT}(t,\ve{r})= U_{cT}\left[\ve{R}_c(t)\right]
+U_{fT}\left[\ve{R}_f(t)\right]-\frac{Z_P Z_T e^2}{R(t)},
\eeqn{tdseV}
where $Z_P$ and $Z_T$ are the atomic numbers of the projectile and the target,
respectively.

The TD equation \eq{tdseeq} has to be solved for all possible trajectories
with the boundary condition that the projectile is initially in its
ground state,
\beq
\Psi^{\rm TD}(t\rightarrow-\infty,b,\ve{r})=\phi_0(\ve{r}).
\eeqn{tdseboundary}
This is performed numerically by applying iteratively an approximation
of the time-evolution operator to the initial wave function
\cite{kido-tdse,esbensen-tdse,typel-tdse,capel-tdse}.
We use the algorithm detailed in Refs.~\cite{esbensen-tdse,esbensen09}.
At the end of the calculation,
a breakup probability can be extracted for
each trajectory by projecting the final wave function on the
positive-energy eigenstates of $H_0$,
\beq
\frac{d P_{\rm bu}}{dk}(b)\propto\sum_{ljIM}|\langle\phi_k^{ljIM}|\Psi^{\rm TD}(t\rightarrow+\infty,b) \rangle|^2.
\eeqn{tdsePbu}
Breakup observables can be calculated from these probabilities by
proper integration over $b$ \cite{capel-tdse}.
Since these observables are obtained by summation over breakup
\emph{probabilities} and not over breakup \emph{amplitudes},
the time-dependent technique cannot account for quantum interferences
between different trajectories.
We will see in \Sec{results} the effects of such interferences.

\subsection{Dynamical eikonal approximation}\label{dea}
More recently, the DEA has been developed from the comparison between
the time dependent model and the eikonal approximation \cite{BCG05,capel-dea}.
It relies on the eikonal factorization of the
three-body wave function \eq{cdccwf0} \cite{Glauber}
\beq
\Psi^{\rm DEA}(\ve{R},\ve{r})=
e^{i K_0 Z}\widehat\Psi(\ve{R},\ve{r}).
\eeqn{deawf}
At sufficiently high energy, the deviation from the initial plane wave $e^{i K_0 Z}$ of the $P$-$T$ relative motion 
is expected to be small.
The dependence on $\ve{R}$ of $\widehat\Psi$ is thus
expected to be smooth. This enables us to neglect its second-order
derivative in \ve{R} with respect to its first-order derivative
\beq
\Delta_{\ve{R}}\widehat\Psi(\ve{R},\ve{r})\ll K_0 \partial/\partial Z\widehat\Psi(\ve{R},\ve{r}).
\eeqn{deaapp}
Therefore, introducing the factorization \eq{deawf} into the
three-body \Sch equation \eq{3beq}, leads to the DEA equation \cite{capel-dea}
\beq
\lefteqn{i\frac{\hbar^2K_0}{\mu_{PT}}\frac{\partial}{\partial Z}
\widehat\Psi(Z,\ve{b},\ve{r})=}\nonumber \\
& & \left[(H_0-E_0)+U_{cT}(\ve{R}_c)+U_{fT}(\ve{R}_f)\right]
\widehat{\Psi}(Z,\ve{b},\ve{r}),
\eeqn{deaeq}
where the dependence of the wave function on the
longitudinal $Z$ and transverse $\ve{b}$ parts of the projectile-target
coordinate $\ve{R}$ has been made explicit (see \fig{f0}).

The DEA equation \eq{deaeq} is mathematically equivalent to a time-dependent
\Sch equation for a straight-line trajectory [see \Eq{tdseeq}].
It can therefore be solved using similar numerical techniques as in the
time-dependent model \cite{kido-tdse,esbensen-tdse,typel-tdse,capel-tdse}.
As explained in \Ref{capel-dea}, we use the algorithm detailed in \Ref{capel-tdse}.
The solution is obtained for each transverse component $\ve{b}$
of the $P$-$T$ coordinate with the boundary condition
\beq
\widehat{\Psi}(Z\rightarrow-\infty,\ve{b},\ve{r})=\phi_0(\ve{r}).
\eeqn{deaboundary}
Breakup amplitudes can then be extracted from the wave function
\beq
S_{\rm bu}^{ljIM}(k,\ve{b})\propto\langle\phi_k^{ljIM}|\widehat\Psi(Z\rightarrow+\infty,\ve{b})\rangle.
\eeqn{skljm}
Since no semiclassical approximation has been made
to derive \Eq{deaeq}, the coordinates $Z$ and $\ve{b}$ are quantal variables.
This enables us to take into account interferences between
trajectories.
This is noticeable by the fact that breakup observables are obtained in the DEA
by integrating \emph{amplitudes} \eq{skljm} over $\ve{b}$ and
not breakup probabilities as in the time dependent method of \Sec{tdse}.
The domain of validity of the approximation \eq{deaapp} remains to be
tested by comparison to the other reaction models.

\section{Model inputs}\label{inputs}
In such a comparison, it is important to ensure that the inputs of all models are consistent.
In this section, we summarize the various parameters that have been considered
in the calculations presented in \Sec{results}.
More details can be found in the Addendum given as supplemental material.

All masses are calculated as mass number times
the nucleon mass $m_N=931.5$~MeV/$c^2$.
The effective potentials simulating the interactions between \ex{14}C, $n$,
and Pb are chosen identical in all three models.
For the \ex{14}C-$n$ potential we take a Woods-Saxon central form factor
(with depth $V_{ws}=63.023$~MeV, radius $R_0=2.651$~fm, and diffuseness $a=0.6$~fm)
plus a spin-orbit term (with depth $V_{so}=23.761$~MeVfm\ex{2},
and the same radius and diffuseness as the central term).
This potential reproduces the two bound states of
$^{15}$C: the $1/2^+$ ground state in the $1s_{1/2}$
partial wave at the experimental energy $E_{1s_{1/2}}=-1.218$~MeV and
the excited $5/2^+$ excited state as a $0d_{5/2}$ state at $E_{0d_{5/2}}=-0.478$~MeV.

The interactions between the Pb target and the projectile constituents are
simulated by optical potentials chosen from the literature.
The Becchetti and Greenlees parametrization \cite{BG69} is used for
the $n$-Pb potential. Since no \ex{14}C-Pb potential is available,
we use, at $68$~MeV/nucleon, a potential reproducing the elastic scattering
of \ex{16}O on Pb at 94 MeV/nucleon \cite{Rou88} and, at $20$~MeV/nucleon,
a potential fitted to the elastic scattering of \ex{16}O on Pb at 312.6~MeV
(potential I3 of \Ref{Olm78}).
At both energies, the radius of the potential is scaled by $(14^{1/3}+208^{1/3})/(16^{1/3}+208^{1/3})$
to correct for the difference between the sizes of \ex{14}C and \ex{16}O.

Convergence is an important part of the study and therefore was thoroughly tested for all cases.
The parameter sets quoted in the Addendum provided as supplemental material
ensure an accuracy of at least 4\% in the energy and angular distributions
for all three models.

\section{Results}
\label{results}

In this section we present the comparison for the breakup of $^{15}$C on $^{208}$Pb at two different
beam energies: the first (68~MeV/nucleon) corresponds to an energy typical of fragmentation facilities
for which data already exist \cite{c15-riken}, and the second (20~MeV/nucleon) serves as an example
of the energies that will be available in facilities such as SPIRAL2 and FRIB.
We present the breakup cross section as a function of either the \ex{14}C-$n$ relative energy $E$
or the scattering angle $\theta$ of the \ex{14}C-$n$ center of mass system.

\begin{figure}
\center
\includegraphics[width=8cm]{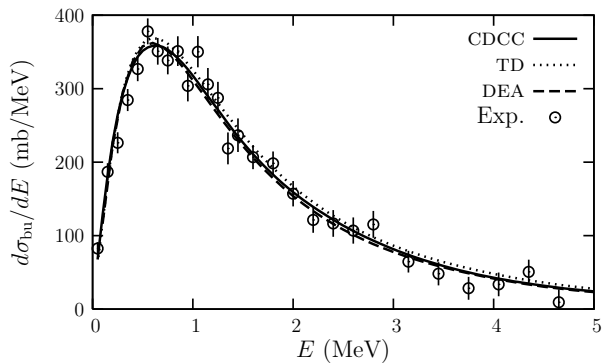}
\caption{Distribution for the breakup of \ex{15}C on Pb at
68~MeV/nucleon as a function of the \ex{14}C-$n$ relative energy.
Comparison of three models: CDCC (solid), TD (dotted), and DEA (dashed).
Experimental data from \Ref{c15-riken}.}\label{f1}
\end{figure}

\begin{figure}
\center
\includegraphics[width=8cm]{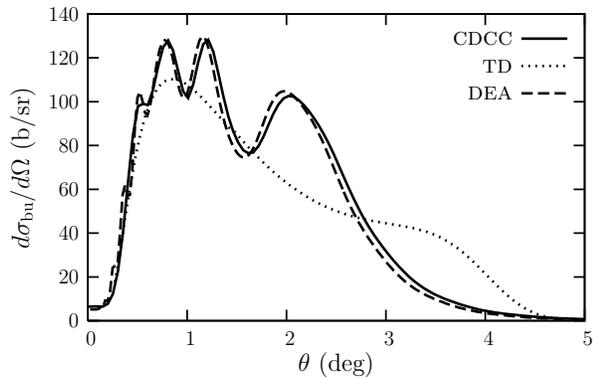}
\caption{Breakup cross section for \ex{15}C on Pb at 68~MeV/nucleon
as a function of the scattering angle of the \ex{14}C-$n$ center of mass.
}\label{f2}
\end{figure}

The results for \ex{15}C on Pb at 68~MeV/nucleon are presented in Figs.~\ref{f1} and \ref{f2}.
All three models predict nearly identical energy distributions (see \fig{f1}):
they differ by only 1--3\% at the peak.
They are also in excellent agreement with the RIKEN data \cite{c15-riken},
validating the reaction theory and the assumed single-particle nature of \ex{15}C.
The aim of this analysis being to compare theories to each other,
the theoretical cross sections have not been folded with the experimental
energy resolution. Such a folding does not affect much the agreement between
theory and experiment in the present case.

The angular distributions are shown in Fig.~\ref{f2}.
The two quantal models, CDCC and DEA, agree very well with each other.
In particular they exhibit similar diffraction patterns.
The TD model does not exhibit any diffractive pattern. This diffraction pattern
is a quantal effect corresponding to interferences between trajectories,
an effect excluded in the semiclassical approximation.
Nevertheless, the TD calculation reproduces the general
trend of the angular distribution at forward angles. This explains why,
once integrated over the scattering angle, it produces a cross section
nearly identical to the quantal models.
Although DEA provides a good approximation to CDCC,
a slight shift of about 3\% in $\theta$
is observed between both oscillatory patterns.
However, at such beam energy,
this small discrepancy is negligible compared to the uncertainties in
the optical potentials.

\begin{figure}
\center
\includegraphics[width=8cm]{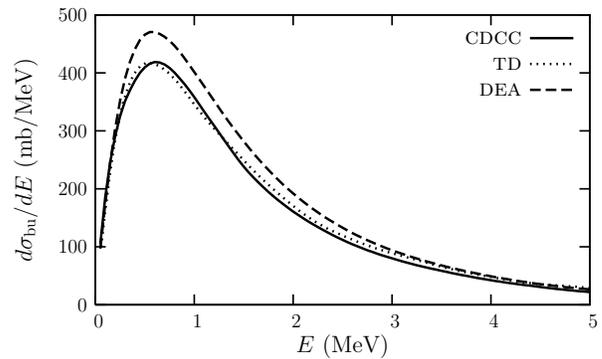}
\caption{Distribution for the breakup of \ex{15}C on Pb at
20~MeV/nucleon as a function of the \ex{14}C-$n$ relative energy.
}\label{f3}
\end{figure}

\begin{figure}
\center
\includegraphics[width=8cm]{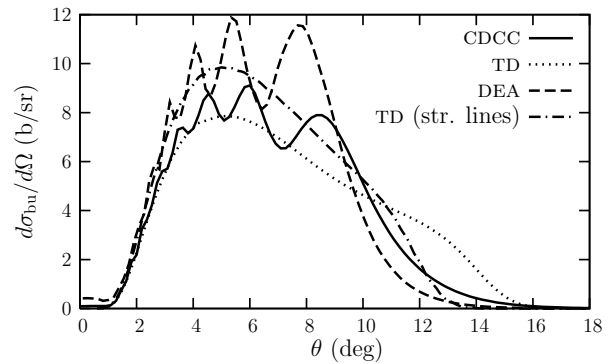}
\caption{Breakup cross section for \ex{15}C on Pb at 20~MeV/nucleon
as a function of the scattering angle of the \ex{14}C-$n$ center of mass.
In addition to CDCC, TD, and DEA results, a TD calculation using straight-line
trajectories (dash-dotted) is shown.
}\label{f4}
\end{figure}

Next we analyze the breakup of \ex{15}C on Pb at 20~MeV/nucleon. The energy distribution is
displayed in Fig.~\ref{f3} and the angular distribution in Fig.~\ref{f4}.
If we first focus on the comparison of TD and CDCC models,
excellent agreement in the energy distribution is found
(a mere 1\% difference at the peak).
At this energy too, the semiclassical approximation fails at reproducing the correct
diffraction pattern seen in the CDCC angular distribution, but, as seen at higher
energy, the general trend of the cross section is well approximated by
the TD model at forward angles.
These results show that the TD model provides accurate breakup observables
integrated over the scattering angle 
even at energies below the range of validity
mentioned by Alder and Winther \cite{AW75}.
Because of its semiclassical approximation, the TD model cannot account for
quantal interferences in the angular distributions. Nevertheless,
it produces a qualitative estimate of the behavior of such distributions
at forward angles.

The DEA energy distribution does not agree with the other two models
at 20~MeV/nucleon; it is about 10\% too high at the peak.
Due to its quantal nature, DEA does exhibit a diffraction pattern in
the angular distribution, but the small discrepancy with CDCC found at
68~MeV/nucleon is now significantly increased
as the DEA angular distribution peaks at more forward angles.
The shift reaches here 10\% in $\theta$.
These results suggest that the difference observed between DEA
and the other two models at low energy comes primarily from
the lack of Coulomb deflection in DEA:
Relying on the eikonal approximation, the DEA assumes that the
incoming plane-wave motion of the projectile is not much perturbed by
its interaction with the target \eq{deaapp}.
The DEA thus forces the projectile straight ahead into the high-field
zone of the target, leading to a larger breakup cross section and a more
forward angular distribution.
On the contrary, the usual TD approach, being based on Coulomb trajectories,
naturally  includes the Coulomb deflection and hence reproduces CDCC
calculations fairly well.

\begin{figure}
\center
\includegraphics[width=8cm]{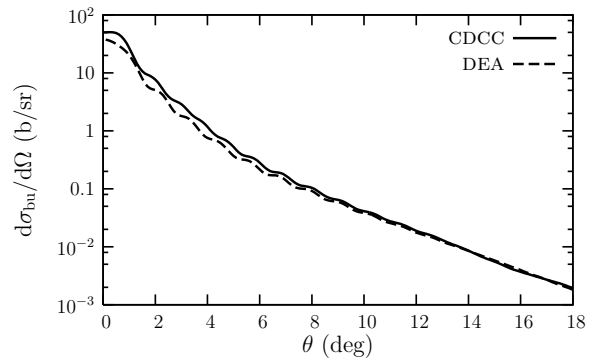}
\caption{Angular distributions for a hypothetical purely nuclear collision
between \ex{15}C on Pb at 20~MeV/nucleon.
DEA is compared to CDCC.
}\label{f5}
\end{figure}

To test this hypothesis, we first repeat the time-dependent calculation
using straight-line trajectories instead of hyperbolas
(dash-dotted line in \fig{f4}).
Of course, this TD calculation does not exhibit any diffraction pattern.
However, it provides a fair approximation of the general trend
of the DEA angular distribution in the same way
the usual TD calculation follows the CDCC one
(compare the dotted and solid lines).
This result was to be expected as the DEA equation \eq{deaeq} is
mathematically equivalent to a time-dependent \Sch equation with
straight-line trajectories (see \Sec{dea}).
It nevertheless confirms the significance of Coulomb deflection in the reaction process.
Second, we compare DEA to CDCC in a purely nuclear calculation,
i.e. setting $Z_T=0$.
The corresponding angular distributions are shown in \fig{f5}.
At large angles, both calculations are nearly identical.
At forward angles, however, DEA \emph{underestimates} CDCC and exhibits an
oscillatory pattern shifted to \emph{larger} angles.
This difference with the Coulomb-dominated reaction is not very surprising
as the nuclear interaction, being mostly attractive, tends to deflect
the projectile within the high-field zone of the target.
This very stringent test indicates that Coulomb deflection
is not the only reason for the discrepancies observed in
Figs.~\ref{f3} and \ref{f4} and that other effects,
such as nuclear deflection and/or couplings between various 
impact parameters $b$, are also significant.

These results confirm that at low energies, the approximation
\eq{deaapp} can no longer be performed as it suppresses part of the deflection
of the projectile by the target, and/or some coupling effects between
different $b$s.
A correction of the DEA that could account for the Coulomb deflection
would most likely provide a better description of
Coulomb-dominated reactions at low-energy.

To make sure that
the qualitative features of our analysis do not depend on the particular
choice of the core-target interaction, we have repeated the
Coulomb-breakup calculations at
20~MeV/nucleon using the optical potentials used at 68~MeV/nucleon. 
As expected, the cross sections are sensitive to the parametrization of
these potentials.
The 68~MeV/nucleon potentials changes the energy distribution by
2--5\% and reduces the amplitude of the oscillations of the diffraction pattern
of the angular distribution.
It also shifts that pattern by about 3\% to larger angles.
Nevertheless, the qualitative differences between the three models
remain very similar.

The difference between CDCC and DEA seems to evolve smoothly when reducing
the beam energy. At 40~MeV/nucleon, the energy distributions differ
by a mere 5\% at the maximum and the shift in the angular distributions
is also about 5\% in $\theta$.

\section{Conclusions}\label{conclusions}
In this study, we perform a comparison of non-perturbative
models of reactions involving loosely-bound nuclei.
We compare the continuum-discretized coupled channel method (CDCC),
the time dependent approach (TD) and the dynamical eikonal approximation (DEA)
for the dissociation of a one-neutron halo nucleus on a heavy target.
Starting from exactly the same
three-body Hamiltonian, we calculate the energy distribution and angular distribution following
the breakup of \ex{15}C on Pb at 68~MeV/nucleon and 20~MeV/nucleon in all three frameworks.

Our results show that for angle-integrated observables,
TD works well and can be safely used in the analysis of data obtained at both
intermediate-energy and low-energy facilities, i.e., on an energy range
much larger than suggested in the original
semiclassical approximation of Alder and Wither \cite{AW75}.
However, due to its classical treatment of trajectories, TD cannot
account for the diffraction pattern seen in the angular distributions.
It provides only the general trend of these cross sections
at forward angles.

The DEA approach is able to accurately reproduce the CDCC angular and energy distributions at 68~MeV/nucleon
and therefore provides a computationally-efficient alternative to CDCC without sacrificing accuracy.
In contrast, at the lower beam energy, both energy and angular distributions in DEA
cannot reproduce the CDCC results.
DEA overestimates the energy distribution by 10\% and,
although the DEA angular distribution exhibits a diffraction pattern similar to that of CDCC,
this pattern is shifted to more forward angles by about 10\%.
The primary cause of these discrepancies is the approximation
\eq{deaapp} made in DEA. For Coulomb-dominated reactions, it amounts
mostly to the absence of Coulomb deflection in that model.
Thanks to the present analysis we now understand how to
remedy the problem so that the domain of validity of DEA
can be partially extended to the lower energies.

Although valid at all energies, CDCC is a reaction model that requires
significant computational power.
Our analysis shows that, depending on the beam energy and/or the
observable considered, it can be reliably replaced by the TD model
or the DEA.
Since the DEA and TD techniques are computationally less expensive,
these could allow for improving the description of the
projectile in reaction models at a reasonable cost.

The present study corresponds to the first comparison of non-perturbative
breakup models at intermediate energies.
It quantitatively shows for which observables and energies the models
agree and in which conditions their predictions should be
considered with caution.
This provides for the first time the range of validity of the three models.
The projectile description being quite general, these results can be 
extended to other neutral loosel-bound systems with confidence.
Note that they cannot be readily extended to charged systems as
the mechanism of the Coulomb dissociation of proton-halo nuclei
differs from that of neutron halos:
the former involves more significant $E2$ transitions and
stronger higher-order dynamical effects than the latter
\cite{EBS05,Oga06,GCB07,EB02B,capel-cc}.

Being focused on the comparison between three reaction models,
the present study has been performed within the framework of
non-relativistic quantum mechanics.
However, relativistic effects may start to play a role at energies
around 100~MeV/nucleon \cite{Ber05}. Our conclusions should therefore not
readily be extended to such energies.
A detailed analysis of the effect of relativity in breakup reactions
is planned in the near future.

\begin{acknowledgments}
We thank I.J.~Thompson for his help in running {\sc fresco} and the support
of the High-Performance Computer Center of MSU while performing our calculations.
P.C. and F.M.N. were supported by the National Science Foundation grant PHY-1068571
and the Department of Energy under contract DE-FG52-08NA28552 and DE-SC0004087.
H.E.  was supported by the U.S. Department of Energy, Office of Nuclear Physics,
under contract No. DE-AC02-06CH11357.
\end{acknowledgments}

\end{document}